\documentclass[11pt,english,pra]{revtex4}
\usepackage[T1]{fontenc}
\usepackage[latin9]{inputenc}
\usepackage[letterpaper]{geometry}
\usepackage{amsthm}
\usepackage{amsmath}
\usepackage{amssymb}

\makeatletter
\@ifundefined{textcolor}{}
{%
 \definecolor{BLACK}{gray}{0}
 \definecolor{WHITE}{gray}{1}
 \definecolor{RED}{rgb}{1,0,0}
 \definecolor{GREEN}{rgb}{0,1,0}
 \definecolor{BLUE}{rgb}{0,0,1}
 \definecolor{CYAN}{cmyk}{1,0,0,0}
 \definecolor{MAGENTA}{cmyk}{0,1,0,0}
 \definecolor{YELLOW}{cmyk}{0,0,1,0}
 }
\theoremstyle{plain}

\makeatother

  \newtheorem{cor}{Corollary} 
\newtheorem{prop}{Proposition} 
\newtheorem{example}{Example} 
\makeatother

\usepackage{babel}

\makeatother

\usepackage{babel}

\begin{document}

\title{LOCC distinguishability of unilaterally transformable quantum states }

\author{Somshubhro Bandyopadhyay }

\altaffiliation{Department of Physics and Center for Astroparticle Physics and Space Science, Bose Institute, Kolkata 700091, India}

\email{som@bosemain.boseinst.ac.in }

\author{Sibasish Ghosh}

\altaffiliation{Optics and Quantum Information Group, The Institute of Mathematical Sciences, C. I. T. Campus, Taramani, Chennai 600113, India }

\email{sibasish@imsc.res.in}

\author{Guruprasad Kar}

\altaffiliation{Physics and Applied Mathematics Unit, Indian Statistical Institute, 203 B. T. Road, Kolkata 700108, India }

\email{gkar@isical.ac.in}
\begin{abstract}
We consider the question of perfect local distinguishability of mutually
orthogonal bipartite quantum states, with the property that every
state can be specified by a unitary operator acting on the local Hilbert
space of Bob. We show that if the states can be exactly discriminated
by one-way LOCC where Alice goes first, then the unitary operators
can also be perfectly distinguished by an orthogonal measurement on
Bob's Hilbert space. We give examples of sets of $N\leq d$ maximally
entangled states in $d\otimes d$ for $d=4,5,6$ that are not perfectly
distinguishable by one-way LOCC. Interestingly for $d=5,6$ our examples
consist of four and five states respectively. We conjecture that these
states cannot be perfectly discriminated by two-way LOCC. 
\end{abstract}
\maketitle

\section{Introduction }

The question of local discrimination of orthogonal quantum states
has received considerable attention in recent years \cite{Bennett-I-99,Bennett-II-99,Chefles2004,Chen+Li,ChenYang2002,Cohen2007,Duan2007,DuanFeng2007,Fan-2004,Ghosh-2001,Ghosh-2004,Hayashi-etal-2006,Horodecki-2003,Nathanson-2005,Virmanietal2001,Walgate-2000,Walgate-2002,Watrous-2005}.
In the bipartite setting, Alice and Bob share a quantum system prepared
in one of a known set of mutually orthogonal quantum states. Their
goal is to determine the state in which the quantum system was prepared
using only local operations and classical communication (LOCC). In
some cases it is possible to identify the state without error while
in some others it is not by LOCC alone. For example, while any two
orthogonal pure states can be perfectly distinguished by LOCC \cite{Walgate-2000},
a complete orthogonal basis of entangled states is locally indistinguishable
\cite{Ghosh-2002,Ghosh-2004,Horodecki-2003}. The nonlocal nature
of quantum information is therefore revealed when a set of orthogonal
states of a composite quantum system cannot be reliably identified
by LOCC. This has been particularly useful to explore quantum nonlocality
and its relationship with entanglement \cite{Bennett-I-99,Bennett-II-99,Horodecki-2003,Walgate-2002},
and has also found practical applications in quantum cryptography
primitives like secret sharing, and data hiding \cite{DiVincenzo2002,Terhal2001,Eggeling2002,Markham-Sanders-2008}. 

The fundamental result of Walgate et al shows that it is always possible
to perfectly discriminate any two orthogonal quantum states by LOCC
regardless of their dimension, multipartite structure and entanglement
\cite{Walgate-2000}. As it turns out, quite remarkably, perfect discrimination
of more than two orthogonal states is not always possible. Examples
include, any three orthogonal entangled states in $2\otimes2$, two
maximally entangled states and a product state in $2\otimes2$ and
so on \cite{Ghosh-2002}. When perfect discrimination is not possible,
one may distinguish the states conclusively or unambigously \cite{Chefles2004,ChenYang2002,Virmanietal2001},
where the unknown state is reliably identified with probability less
than unity. A necessary and sufficient condition for unambigious discrimination
of quantum states, not necessarily orthogonal was obtained by Chefles
\cite{Chefles2004}. Recently, Bandyopadhyay and Walgate has shown
that for any set of three states conclusive identification is always
possible \cite{Bandyo+Walgate-2009}. In the worst case scenario,
only one member of the set, and not all, can be correctly identified,
albeit with a non-zero probability. 

Interestingly, the maximally entangled basis (Bell basis) in $2\otimes2$
\cite{Ghosh-2001}, or a complete orthogonal entangled basis in $n\otimes m$
\cite{Horodecki-2003} are not even conclusively distnguishable, in
which case we say that the sets are completely indistinguishable.
Note that if an orthogonal set contains at least one product state,
one can always distinguish the set conclusively. Therefore, all members
of a completely indistinguishable set must necessarily be entangled.

The present work is motivated by the results on local distinguishability
of orthogonal maximally entangled states \cite{Ghosh-2001,Ghosh-2004,Fan-2004,Nathanson-2005},
and in particular those that put an upper bound on the number of states
that can be perfectly distinguished by LOCC \cite{Ghosh-2004,Nathanson-2005}.
For example, it was first observed in \cite{Ghosh-2004} that no more
than $d$ maximally entangled states in $d\otimes d$ can be perfectly
distinguished provided the states were chosen from the Bell basis.
This was soon followed by a more general result establishing this
bound for any set of maximally entangled states in $d\otimes d$ \cite{Nathanson-2005}. 

It is therefore natural to ask whether any $N$ orthogonal maximally
entangled states in $d\otimes d$ can be perfectly distinguished by
means of a LOCC protocol if $N\leq d$. The general answer is not
yet known except in dimensions $2\otimes2$ \cite{Walgate-2000} and
$3\otimes3$ \cite{Nathanson-2005}. In $2\otimes2$ the answer follows
as a corollary of the more general result that any two orthogonal
quantum states of a composite quantum system can be reliably distinguished
\cite{Walgate-2000}. In \cite{Nathanson-2005} a constructive proof
was given to show that any three orthogonal maximally entangled states
in $3\otimes3$ can be perfectly distinguished by LOCC. It is worth
noting that in both \cite{Walgate-2000} and \cite{Nathanson-2005}
the maximally entangled states could be perfectly distinguished by
one-way LOCC. Indeed, for almost all known sets of bipartite orthogonal
states that are perfectly LOCC distinguishable, one-way protocols
are sufficient. A notable exception to this can be found in \cite{Bennett-I-99}
where it was shown that two-way LOCC is required to distinguish subsets
of a locally indistinguishable orthogonal basis of $3\otimes3$.

\section{Formulation of the Problem and Results}

In this work we consider the question of perfect LOCC distinguishability
of bipartite orthogonal quantum states $|\psi_{1}\rangle,|\psi_{2}\rangle,...,|\psi_{N}\rangle\in\mathcal{H}_{A}\otimes\mathcal{H}_{B}$
with the property \begin{equation}
|\psi_{i}\rangle=\left(I\otimes U_{i}\right)|\psi\rangle,\label{def-states}\end{equation}
$i=1,...,N$ for $U_{i}$ unitary. Eq.$\,$(\ref{def-states}) is
equivalent to the fact that Bob alone can transform $|\psi_{i}\rangle$
into $|\psi_{j}\rangle$ for every pair $\left(i,j\right)$. These
states are known not to be perfectly distinguishable by LOCC if $N>\dim\mathcal{H}_{B}$
\cite{Nathanson-2005}. Therefore, the only case of interest is $N\leq\dim\mathcal{H}_{B}$.
Clearly, for a given $|\psi_{1}\rangle$, the states defined by (\ref{def-states})
are completely specified by the set of unitary operators $\left\{ U_{1},U_{2},...,U_{N}\right\} $
on $\mathcal{H}_{B}$. Let us point out that the maximally entangled
states form a subset of the class of sets defined by (\ref{def-states}). 

The main result of this paper lies in showing that one-way LOCC distinguishability
of the states (\ref{def-states}) can be completely characterized
by distinguishability of the unitary operators $\left\{ U_{1},U_{2},...,U_{N}\right\} $
acting on Bob's Hilbert space. Before we proceed let us first explain
what we mean by distinguishing unitary operators. 

A given set of unitary operators $\left\{ U_{1},U_{2},...,U_{n}\right\} $
acting on some Hilbert space $\mathcal{H}$ is said to perfectly distinguishable
in $\mathcal{H}$ if there exists a vector $|\eta\rangle\in\mathcal{H}$
such that\begin{equation}
\langle\eta|U_{i}^{\dagger}U_{j}|\eta\rangle=\delta_{ij}\label{ortho-eta}\end{equation}
for all $1\leq i,j\leq n$. It could so happen that such a vector
$|\eta\rangle$ does not exist. This however, does not mean that the
unitary operators cannot be reliably distinguished because it may
be possible to discriminate them exactly in a locally extended tensor
product space. 

A set of unitary operators $\left\{ U_{1},U_{2},...,U_{n}\right\} $
on $\mathcal{H}$ are perfectly distinguishable in an extended tensor
product space $\mathcal{H}^{'}\otimes\mathcal{H}$ if there exists
a vector $|\zeta\rangle\in\mathcal{H}^{'}\otimes\mathcal{H}$ such
that\begin{equation}
\langle\zeta|\left(I\otimes U_{i}^{\dagger}U_{j}\right)|\zeta\rangle=\delta_{ij}\label{ortho-zeta}\end{equation}
for all $1\leq i,j\leq n$. The above equation simply reflects the
orthogonality condition for the vectors $\left(I\otimes U_{i}\right)|\zeta\rangle\in\mathcal{H}^{'}\otimes\mathcal{H}$
for $i=1,...,n$. Notice that if (\ref{ortho-eta}) holds then so
does (\ref{ortho-zeta}) trivially. The converse however, is not generally
true. In this paper we are particularly interested in those unitary
operators that cannot be perfectly distinguished in the Hilbert space
they act upon but instead they can be distinguished in an extended
tensor product space. Notice that as far as distinguishing a set of
unitary operators are concerned, the question of LOCC doesn't arise
for obvious reasons. The tensor product extension can be done locally
by bringing in an ancilla. 

We can now state our results.

\begin{prop} Let $\left\{ U_{1},U_{2},...,U_{n}\right\} $ be a set
of unitary operators on Hilbert space $\mathcal{H}$, where, $n\leq\dim\mathcal{H}$.
If the unitary operators can be perfectly distinguished only in an
extended tensor product Hilbert space $\mathcal{H}^{'}\otimes\mathcal{H}$,
then there exists a set of orthogonal states\begin{equation}
|\psi_{i}\rangle=\left(I\otimes U_{i}\right)|\psi\rangle\label{def-state-1}\end{equation}
in $\mathcal{H}^{'}\otimes\mathcal{H}$ for some vector $|\psi\rangle\in\mathcal{H}^{'}\otimes\mathcal{H}$
and $i=1,...,n$. The set of states $\left\{ |\psi_{1}\rangle,|\psi_{2}\rangle,...,|\psi_{n}\rangle\right\} $
is not perfectly distinguishable by one-way LOCC in the direction
$\mathcal{H}^{'}\rightarrow\mathcal{H}$ where the class of LOCC operations
are defined with respect to the tensor product space $\mathcal{H}^{'}\otimes\mathcal{H}$.
\end{prop}

An immediate consequence of this result is that if the orthogonal
states defined by (\ref{def-states}) are perfectly distinguishable
by one-way LOCC where Alice goes first, then the local unitary operators
$U_{1},U_{2},...,U_{N}$ can also be perfectly distinguished in $\mathcal{H}_{B}$. 

\begin{cor} Consider a set of mutually orthogonal vectors $\left\{ |\psi_{i}\rangle,|\psi_{2}\rangle,...,|\psi_{N}\rangle\right\} $
in $\mathcal{H}_{A}\otimes\mathcal{H}_{B}$ with the property that
for every $i$, $|\psi_{i}\rangle=\left(I\otimes U_{i}\right)|\psi_{1}\rangle$
for $U_{i}$ unitary. Futhermore $N\leq\dim\mathcal{H}_{B}$. If the
vectors are perfectly distinguishable by one-way LOCC in the direction
$A\rightarrow B$ then there exists at least one vector $|\phi\rangle\in\mathcal{H_{B}}$
such that for all $k,l,$ with $1\leq k,l\leq N$, $\langle\phi|U_{k}^{\dagger}U_{l}|\phi\rangle=\delta_{kl}$.
\end{cor} 

Observe that the necessary condition is nontrivial and interesting
only if $N\leq\dim\mathcal{H}_{B}$. Otherwise it is trivially violated.
Interestingly if the states are in $2\otimes d$ then the above condition
holds for all two-way LOCC protocols initiated by Alice.

\begin{cor} Consider a set of mutually orthogonal vectors $\left\{ |\psi_{i}\rangle,|\psi_{2}\rangle,...,|\psi_{N}\rangle\right\} \in\mathcal{H}_{A}\otimes\mathcal{H}_{B}$,
$\dim\mathcal{H}_{A}=2$, $\dim\mathcal{H}_{B}\geq2$, with the property
that $|\psi_{i}\rangle=\left(I\otimes U_{i}\right)|\psi_{1}\rangle$,
$i=1,...,N$ for $U_{i}$ unitary. Furthermore $N\leq\dim\mathcal{H}_{B}$.
If the set is perfectly distinguishable by LOCC when Alice goes first,
then there exists at least one vector $|\phi\rangle\in\mathcal{H_{B}}$
such that for any $k,l\in\left\{ 1,2,...,N\right\} $ , $\langle\phi|U_{k}^{\dagger}U_{l}|\phi\rangle=\delta_{kl}$.
\end{cor}

We apply our results to the case of distinguishing maximally entangled
states. We notice that similar property as in (\ref{def-states})
holds for maximally entangled states as well. That is, if $|\Psi\rangle$
is a maximally entangled state of $d\otimes d$ then it can be written
in terms of the standard maximally entangled state \begin{equation}
|\Phi^{+}\rangle=\frac{1}{\sqrt{d}}\sum_{j=0}^{d-1}|j\rangle|j\rangle,\label{std-max-state}\end{equation}
in the following way:\begin{align}
|\Psi\rangle & =\left(I\otimes U\right)|\Phi^{+}\rangle\label{gen-max-state}\\
 & =\left(U^{T}\otimes I\right)|\Phi^{+}\rangle\label{gen-max-2}\end{align}
where, $U$ is unitary. The following result makes explicit use of
the equations (\ref{gen-max-state}) and (\ref{gen-max-2}) for one-way
LOCC in the directions $A\rightarrow B$ and $B\rightarrow A$ respectively. 

\begin{cor} Consider a set of maximally entangled vectors $\left\{ |\Psi_{1}\rangle,|\Psi_{2}\rangle,...,|\Psi_{N}\rangle\right\} $
in $\mathcal{H}_{A}\otimes\mathcal{H}_{B}$ where, $N\leq\dim\mathcal{H}_{A}=\dim\mathcal{H}_{B}=d$,
with $|\Psi_{i}\rangle=\left(I\otimes U_{i}\right)|\Phi^{+}\rangle$.
If the set is perfectly distinguishable by one-way LOCC in the direction
$A\rightarrow B$, then there exists at least one vector $|\phi\rangle\in\mathcal{H}_{B}$
such that, $\langle\phi|U_{k}^{\dagger}U_{l}|\phi\rangle=\delta_{kl}$
for $1\leq k,l\leq N$. On the other hand if the set is perfectly
distinguishable by one-way LOCC in the direction $B\rightarrow A$,
then there exists at least one vector $|\phi^{'}\rangle\in\mathcal{H}_{A}$
so that $\langle\phi^{'}|V_{k}^{\dagger}V_{l}|\phi^{'}\rangle=\delta_{kl}$
for $1\leq k,l\leq N$, where $V_{k}=U_{k}^{T}$. \end{cor}

We note that the known cases in which a set of maximally entangled
states can be perfectly distinguished by LOCC (these LOCC protocols
are all one-way in the direction $A\rightarrow B$ \cite{Fan-2004,Ghosh-2004,Nathanson-2005}),
the orthogonal measurements on Bob's Hilbert space make explicit use
of vectors $\left\{ |\phi_{m}\rangle\right\} \in\mathcal{H}_{B}$
with the property $\langle\phi_{m}|U_{k}^{\dagger}U_{l}|\phi_{m}\rangle=\delta_{kl}$
for every $m$ and for all $k$ and $l$. 

Given the existing symmetry in maximally entangled states one might
wonder whether there is any difference between the one-way LOCC protocols
{}``Alice goes first'' and {}``Bob goes first''. This is an interesting
question and intuitively it seems that for distinguishing maximally
entangled states this should not be an issue. However we haven't been
able to conclusively prove that this is the case. As noted in Corollary
3, if the states are perfectly distinguishable when Bob goes first
then the orthogonality condition\begin{equation}
\langle\phi^{'}|V_{k}^{\dagger}V_{l}|\phi^{'}\rangle=\delta_{kl}\label{ortho-1}\end{equation}
must hold for all $k$ and $l$ for some $|\phi^{'}\rangle$. Using
the fact that $V_{k}=U_{k}^{T}$ the above equation can also be written
as\begin{equation}
\langle\phi^{'}|U_{k}^{*}U_{l}^{T}|\phi^{'}\rangle=\delta_{kl}\label{ortho-2}\end{equation}
which in turn is equivalent to the condition\begin{equation}
\langle\phi^{*'}|U_{l}U_{k}^{\dagger}|\phi^{*'}\rangle=\delta_{kl}.\label{ortho-3}\end{equation}
 Comparing the above condition with that of one-way LOCC in the direction
$A\rightarrow B$ (as mentioned in Corollary 1) it is not clear if
there is any one-to-one correspondence between the two. So we conclude
that if the maximally entangled states are perfectly distinguishable
by one-way LOCC in the direction $A\rightarrow B$, then they can
also be perfectly distinguished in the opposite direction provided
$\left[U_{k}^{\dagger},U_{l}\right]=0$ for all $k,l=1,...,N$. In
the latter case one can of course choose $|\phi^{'}\rangle=|\phi^{*}\rangle$.

\section{One-way LOCC indistinguishable maximally entangled states}

We now give examples of one-way locally indistinguishable sets of
$N$ orthogonal maximally entangled states in $d\otimes d$, where
$N\leq d$ and $d=4,5,6$. Our examples constitute the following:
(a) a set of four maximally entangled states in $4\otimes4$, (b)
a set of four maximally entangled states in $5\otimes5$, and (c)
a set five maximally entangled states in $6\otimes6$. To show that
these states are locally indistinguishable by all one-way LOCC protocols
it suffices to show (see Corollary 3) that the local unitary operators
(or their transposes) cannot be perfectly distinguished in $\mathcal{H}_{B}$
(or $\mathcal{H}_{A}$). We provide complete proofs for all the examples. 

The maximally entangled states considered in these examples belong
to the family of generalized Bell states. In $d\otimes d,$ $d^{2}$
generalized Bell states written in the standard basis can be expressed
as,\begin{equation}
|\Psi_{nm}^{\left(d\right)}\rangle=\frac{1}{\sqrt{d}}\sum_{j=0}^{d-1}e^{\frac{2\pi ijn}{d}}|j\rangle\otimes|j\oplus_{d}m\rangle\label{Bell-d-general}\end{equation}
 for $n,m=0,1,\cdots,d-1$, where, $j\oplus_{d}m\equiv\left(j+m\right)\mod d$.
The standard maximally entangled state $|\Phi^{+}\rangle$ in $d\otimes d$
is simply $|\Psi_{00}^{\left(d\right)}\rangle=\frac{1}{\sqrt{d}}\sum_{i=0}^{d-1}|i\rangle\otimes|i\rangle$.
These states are related to the standard maximally entangled state
in the following way,\begin{equation}
\left(I\otimes U_{nm}^{(d)}\right)|\Psi_{00}\rangle=|\Psi_{nm}^{\left(d\right)}\rangle\label{Bell-general-2}\end{equation}
 where, \begin{equation}
U_{nm}^{(d)}=\sum_{j=0}^{d-1}e^{\frac{2\pi ijn}{d}}|j\oplus_{d}m\rangle\langle j|\label{U-general}\end{equation}
 are $d\times d$ unitary matrices for $n,m=0,1,\cdots,d-1$ . 

\begin{example} The following four maximally entangled states $|\Psi_{00}^{(4)}\rangle,|\Psi_{11}^{(4)}\rangle,|\Psi_{32}^{(4)}\rangle,|\Psi_{31}^{(4)}\rangle$
in $4\otimes4$ are not perfectly distinguishable by one-way LOCC.
\end{example}

\begin{example} The following four maximally entangled states $|\Psi_{00}^{(5)}\rangle,|\Psi_{01}^{(5)}\rangle,|\Psi_{31}^{(5)}\rangle,|\Psi_{22}^{(5)}\rangle$
in $5\otimes5$ are not perfectly distinguishable by one-way LOCC.
\end{example}

\begin{example} The following five maximally entangled states $|\Psi_{00}^{(6)}\rangle,|\Psi_{01}^{(6)}\rangle,|\Psi_{41}^{(6)}\rangle,|\Psi_{12}^{(6)}\rangle,|\Psi_{33}^{(6)}\rangle$
in $6\otimes6$ are not perfectly distinguishable by one-way LOCC.
\end{example}

\section{Proofs }

\emph{Proof of Proposition 1:} Assume that the unitary operators $U_{1},U_{2},...,U_{n}$
acting on $\mathcal{H}$ can only be distinguished in an extended
tensor product space $\mathcal{H}^{\prime}\otimes\mathcal{H}$. This
implies that there does not exist any vector $|\phi\rangle\in\mathcal{H}$,
such that for all $k,l,$ with $1\leq k,l\leq n$, \begin{equation}
\langle\phi|U_{k}^{\dagger}U_{l}|\phi\rangle=\delta_{kl}.\label{proof-1-eq-1}\end{equation}
We will now show that if the set of states $\left\{ |\psi_{1}\rangle,|\psi_{2}\rangle,...,|\psi_{n}\rangle\right\} $
defined by Eq.$\,$(\ref{def-state-1}) is perfectly distinguishable
by one-way LOCC in the direction $\mathcal{H}^{'}\rightarrow\mathcal{H}$
where the class of LOCC operations are defined with respect to the
tensor product space $\mathcal{H}^{'}\otimes\mathcal{H}$, then there
must exist a vector $|\phi\rangle\in\mathcal{H}$, such that for all
$k,l,$ with $1\leq k,l\leq n$, \begin{equation}
\langle\phi|U_{k}^{\dagger}U_{l}|\phi\rangle=\delta_{kl}.\label{proof-1-eq-1.1}\end{equation}
Suppose that the states $|\psi_{1}\rangle,...,|\psi_{n}\rangle\in\mathcal{H}^{\prime}\otimes\mathcal{H}$
are perfectly distinguishable by one-way LOCC in the direction $\mathcal{H}^{'}\rightarrow\mathcal{H}$.
Let $\mathcal{A}=\left\{ \mathcal{A}_{1},\mathcal{A}_{2},\cdots,\right\} $
be the POVM of the local measurement on $\mathcal{H}^{\prime}$ satisfying
the usual constraints that $\left\{ \mathcal{A}_{i}\right\} $ are
positive operators and $\sum_{i}A_{i}\leq\mathcal{I}_{\mathcal{H}^{\prime}}$.
Associated with the $i^{th}$ outcome, let $\mathcal{B}^{i}=\left\{ \mathcal{B}_{ij}\right\} $
be the POVM of the local measurement on $\mathcal{H}$ satisfying
$\sum_{j}\mathcal{B}_{ij}\leq\mathcal{I}_{\mathcal{H}}$ where $\left\{ \mathcal{B}_{ij}\right\} $
are positive operators. It may be noted that by defining $\mathcal{B}_{i}=\sum_{j}\mathcal{B}_{ij}$,
the collection of positive operators $\left\{ \mathcal{A}_{i}\otimes\mathcal{B}_{i}\right\} $
represents a separable POVM satisfying $\sum_{i}\mathcal{A}_{i}\otimes\mathcal{B}_{i}\leq I_{\mathcal{H}^{\prime}\otimes\mathcal{H}}$. 

Let $\mathcal{A}_{i}=A_{i}^{\dagger}A_{i}$, where $A_{i}$ is the
Kraus element. Subsequent to the $i^{th}$ outcome of the measurement
$\mathcal{A}$, the reduced density matrix on $\mathcal{H}$ (for
the input state $|\psi_{k}\rangle$) is given by\begin{equation}
\sigma_{k,A_{i}}=\mbox{Tr}_{A}\frac{\rho_{k}\mathcal{A}_{i}\otimes I}{\mbox{Tr}\left(\rho_{k}\mathcal{A}_{i}\otimes I\right)},\label{proof-1-eq-2}\end{equation}
where, $\rho_{k}=|\psi_{k}\rangle\langle\psi_{k}|$. Because a measurement
now perfectly distinguishes the set of reduced density matrices $\left\{ \sigma_{k,A_{i}}\in\mathcal{H}\,:k=1,...,n\right\} $,
they must be mutually orthogonal, that is,\begin{equation}
\mbox{Tr}\left(\sigma_{k,A_{i}}\sigma_{l,A_{i}}\right)=0\,:k\neq l\label{proof-1-eq-3}\end{equation}
 Noting that the states we are trying to perfectly distinguish are
of the form,\begin{equation}
|\psi_{k}\rangle=\left(I\otimes U_{k}\right)|\psi_{1}\rangle\label{proof-1-eq-4}\end{equation}
for $k=1,...,n$; the transformed state $|\psi_{k,A_{i}}\rangle$
(unnormalized) post measurement on $\mathcal{H}^{\prime}$ is given
by\begin{equation}
|\psi_{k,A_{i}}\rangle=\left(A_{i}\otimes I\right)\left(I\otimes U_{k}\right)|\psi_{1}\rangle=\left(I\otimes U_{k}\right)|\psi_{1,A_{i}}\rangle\label{proof-1-eq-5}\end{equation}
This in turn implies that the reduced density matrices $\sigma_{k,A_{i}}$
for all $k$, can be expressed in terms of $\sigma_{1,A_{i}}$ as,
\begin{equation}
\sigma_{k,A_{i}}=U_{k}\sigma_{1,A_{i}}U_{k}^{\dagger}\label{proof-1-eq-6}\end{equation}
 Let the spectral decomposition of the density matrix $\sigma_{1.A_{i}}$
be, \begin{equation}
\sum_{p=1}^{r}\lambda_{p}^{i}|\chi_{p}^{i}\rangle\langle\chi_{p}^{i}|\label{proof-1-eq-7}\end{equation}
where, $0<\lambda_{p}^{i}\leq1,\:\sum_{p=1}^{r}\lambda_{p}^{i}=1,$
and $\langle\chi_{p}^{i}|\chi_{q}^{i}\rangle=\delta_{pq}$. Using
the Eqs.$\,$(\ref{proof-1-eq-6}) and (\ref{proof-1-eq-7}) we can
rewrite $\sigma_{k,A_{i}}$ as, \begin{equation}
\sigma_{k,A_{i}}=\sum_{p=1}^{r}\lambda_{p}^{i}U_{k}|\chi_{p}^{i}\rangle\langle\chi_{p}^{i}|U_{k}^{\dagger}\label{proof-1-eq-8}\end{equation}
 We now apply the orthogonality condition:- $\mbox{Tr}\left(\sigma_{k,A_{i}}\sigma_{l,A_{i}}\right)=0$,
if $k\neq l$ to obtain,\begin{equation}
\mbox{Tr}\left(\sigma_{k,A_{i}}\sigma_{l,A_{i}}\right)=\sum_{p}\left(\lambda_{p}^{i}\right)^{2}|\langle\chi_{p}^{i}|U_{k}^{\dagger}U_{l}|\chi_{p}^{i}\rangle|^{2}+\sum_{p\neq q}\lambda_{p}^{i}\lambda_{q}^{i}|\langle\chi_{p}^{i}|U_{k}^{\dagger}U_{l}|\chi_{q}^{i}\rangle|^{2}=0\label{proof-1-eq-9}\end{equation}
from which it follows that every term in the summation must be identically
zero. This is because each term is non-negative (note that $0<\lambda_{p}^{i}\leq1$)
and by adding all the terms we get zero. Moreover, Eq.$\,$(\ref{proof-1-eq-9})
holds for all $k$ and $l$. Therefore for every $p$ we have, \begin{equation}
|\langle\chi_{p}^{i}|U_{k}^{\dagger}U_{l}|\chi_{p}^{i}\rangle|^{2}=0\label{proof-1-eq-10}\end{equation}
from which it follows that there exist vectors $\left\{ |\chi_{p}\rangle,U_{k}|\chi_{p}\rangle\in\mathcal{H}\;:k=2,...,n\right\} $
forming an orthogonal set. This is in contradiction with the fact
that the unitary operators are distinguishable only in an extended
tensor product space. This proves the result. $\square$\\
\textbf{}\\
\textbf{Remark 1: }As noted before Corollary 1 is a direct consequence
of Proposition 1. The result of Corollary 2, however, holds for all
two way LOCC protocols initiated by Alice. The proof is given below.
\\
\emph{}\\
\emph{Proof of Corollary 2:} We assume that the set of vectors
$\left\{ |\psi_{i}\rangle\,:i=1,...,N\right\} \in2\otimes d$ can
be perfectly distinguished by LOCC if Alice goes first. From a result
in \cite{Walgate-2002} it follows that there exists a basis $\left\{ |0\rangle,|1\rangle\right\} $
for Alice such that in that basis,\begin{equation}
|\psi_{i}\rangle=|0\rangle|\chi_{i}^{0}\rangle+|1\rangle|\chi_{i}^{1}\rangle\label{proof-2-eq-1}\end{equation}
where, $\langle\chi_{i}^{0}|\chi_{j}^{0}\rangle=\langle\chi_{i}^{1}|\chi_{j}^{1}\rangle=0$
if $i\neq j$. Using the fact that for every $i$, \begin{equation}
|\psi_{i}\rangle=\left(I\otimes U_{i}\right)|\psi_{1}\rangle,\label{proof-2-eq-2}\end{equation}
where, $U_{i}$ is unitary, (\ref{proof-2-eq-1}) can be rewritten
as \begin{equation}
|\psi_{i}\rangle=|0\rangle U_{i}|\chi_{1}^{0}\rangle+|1\rangle U_{i}|\chi_{1}^{1}\rangle\label{proof-2-eq-3}\end{equation}
where the states $\left\{ U_{i}|\chi_{1}^{x}\rangle\,:x=0,1\,:i=1,...,N\right\} $
satisfy the following orthogonality conditions\begin{equation}
\langle\chi_{1}^{0}|U_{i}^{\dagger}U_{j}|\chi_{1}^{0}\rangle=\langle\chi_{1}^{1}|U_{i}^{\dagger}U_{j}|\chi_{1}^{1}\rangle=0\label{proof-2-eq-4}\end{equation}
if $i\neq j$. This concludes the proof. $\square$\\
\emph{}\\
\emph{Proof of Corollary 3:} We first note that a given set of
orthogonal maximally entangled vectors can be written in the form
of Eq.$\,$(\ref{def-state-1}) by virtue of Eq.$\,$(\ref{gen-max-state}).
Clearly the results of Proposition 1 and Corollary 1 apply for one
way LOCC protocols in the direction $A\rightarrow B$ where the corresponding
unitary operators acting on Bob's Hilbert space are denoted by $U_{1},...,U_{N}$.
On the other hand owing to Eq.$\,$(\ref{gen-max-2}) we know that
the same given set of maximally entangled states can also be defined
by the action of unitary operators $U_{1}^{T},...,U_{N}^{T}$ acting
only on the local Hilbert space of Alice. Thus the results of Proposition
1 and Corollary 1 also apply to one-way LOCC protocols in the direction
$B\rightarrow A$. $\square$

\section{Proofs of the examples}

\emph{Proof of Example 1}. We will show that the states are not perfectly
distinguishable by one-way LOCC in the direction $A\rightarrow B$.
A similar proof can be worked out in the direction $B\rightarrow A$.
We write the states as: $|\Psi_{00}^{(4)}\rangle,|\Psi_{11}^{(4)}\rangle=\left(I\otimes U_{11}^{(4)}\right)|\Psi_{00}^{(4)}\rangle,|\Psi_{32}^{(4)}\rangle=\left(I\otimes U_{32}^{(4)}\right)|\Psi_{00}^{(4)}\rangle,$
and $|\Psi_{31}^{(4)}\rangle=\left(I\otimes U_{31}^{(4)}\right)|\Psi_{00}^{(4)}\rangle$.
From Corollary 3, a necessary condition for these four states to be
perfectly distinguishable by one-way LOCC in the direction $A\rightarrow B$
is that there must exist a vector (normalized) $|\phi\rangle=\sum_{j=0}^{3}\phi_{j}|j\rangle\in\mathcal{H}_{B}$
satisfying the normalization condition \begin{equation}
\sum_{j-0}^{3}|\phi_{j}|^{2}=1\label{example-1:eq:1}\end{equation}
such that the following four vectors $|\phi\rangle,U_{11}^{(4)}|\phi\rangle,U_{32}^{(4)}|\phi\rangle,U_{31}^{(4)}|\phi\rangle$
are pairwise orthogonal. From here on we will omit the superscript
in the unitaries. It is easy to verify that the six unitary operators
$U_{11},U_{31},U_{32},U_{11}^{\dagger}U_{32},U_{11}^{\dagger}U_{31},U_{32}^{\dagger}U_{31}$
are all distinct. We now write the orthogonality conditions:\begin{eqnarray}
\langle\phi|U_{11}|\phi\rangle=\sum_{j-0}^{3}\omega^{j}\phi_{j}\phi_{j\oplus_{4}1}^{*} & = & 0,\label{example-1-eq-2}\\
\langle\phi|U_{31}|\phi\rangle=\sum_{j=0}^{3}\omega^{3j}\phi_{j}\phi_{j\oplus_{4}1}^{*} & = & 0,\label{example-1-eq-3}\\
\langle\phi|U_{32}|\phi\rangle=\sum_{j=0}^{3}\omega^{3j}\phi_{j}\phi_{j\oplus_{4}2}^{*} & = & 0,\label{example-1-eq-4}\\
\langle\phi|U_{11}^{\dagger}U_{32}|\phi\rangle=\sum_{j=0}^{3}\omega^{2j}\phi_{j}\phi_{j\oplus_{4}1}^{*} & = & 0,\label{example-1-eq-5}\\
\langle\phi|U_{11}^{\dagger}U_{31}|\phi\rangle=\sum_{j=0}^{3}\omega^{2j}|\phi_{j}|^{2} & = & 0,\label{example-1-eq-6}\\
\langle\phi|U_{32}^{\dagger}U_{31}|\phi\rangle=\sum_{j=0}^{3}\phi_{j}\phi_{j\oplus_{4}3}^{*} & = & 0,\label{example-1-eq-7}\end{eqnarray}
where all the exponents of $\omega=e^{\frac{2\pi i}{4}}$ are taken
to be numbers addition modulo 4. From Eqs.$\,$(\ref{example-1-eq-2}),
(\ref{example-1-eq-3})and (\ref{example-1-eq-5}) one finds that
the the vector $\left(\phi_{0}^{*}\phi_{1},\phi_{1}^{*}\phi_{2},\phi_{2}^{*}\phi_{3},\phi_{3}^{*}\phi_{0}\right)\in\mathbb{C}^{4}$
is orthogonal to the following three vectors: $\left(1,\omega,\omega^{2},\omega^{3}\right)$,
$\left(1,\omega^{3},\omega^{2},\omega\right)$, and $\left(1,\omega^{2},1,\omega^{2}\right)$.
Therefore, we must have, \begin{equation}
\left(\phi_{0}^{*}\phi_{1},\phi_{1}^{*}\phi_{2},\phi_{2}^{*}\phi_{3},\phi_{3}^{*}\phi_{0}\right)=\lambda\left(1,1,1,1\right)\label{example-1-eq-8}\end{equation}
for some $\lambda\in\mathbb{C}$. We will show that the above equality
cannot be satisfied except when $\phi_{i}=0$ for every $i$ and $\lambda=0$
thereby completing the proof. To show this we need to consider two
cases, namely, $\lambda\neq0$ and $\lambda=0$. 

Case 1 ($\lambda\neq0$): From Eq.$\,$(\ref{example-1-eq-8})), here
we must have, $\forall j$, $\phi_{j}\neq0.$ Thus for $j=0,1,2,3$,
we have the following two relations:\begin{align*}
\phi_{j}^{*}\phi_{j\oplus_{4}2} & =\frac{\lambda^{2}}{|\phi_{j\oplus_{4}1}|^{2}},\\
\phi_{j}^{*}\phi_{j\oplus_{4}3} & =\frac{\lambda^{3}}{|\phi_{j\oplus_{4}1}\phi_{j\oplus_{4}2}|^{2}}.\end{align*}
Then from Eq.$\,$(\ref{example-1-eq-7}) we see that\[
\lambda^{*3}\sum_{j=0}^{3}\frac{1}{|\phi_{j\oplus_{4}1}\phi_{j\oplus_{4}2}|^{2}}=0\]
immediately implying that $\lambda=0$ which is a contradiction. 

Case 2 ($\lambda=0$): Here the nontrivial cases arise only when any
two $\phi_{i}$s are zero and the remaining two are non-zero. It is
simple to verify that a contradiction is always reached. For example,
suppose $\phi_{0}=\phi_{2}=0$ and $\phi_{1}\neq0,\,\phi_{3}\neq0$.
From Eq.$\,$(\ref{example-1-eq-6}) we obtain $|\phi_{1}|^{2}+|\phi_{3}|^{2}=0$,
which immediately implies that $\phi_{1}=\phi_{3}=0$. This therefore
completes the proof. $\square$\\

\emph{Proof of example 2:} We will prove local indistinguishability
in the direction $A\rightarrow B$. A similar proof holds for $B\rightarrow A$
as well. Consider the following four maximally entangled states in
$5\otimes5$:\begin{align*}
|\Psi_{00}\rangle & =\frac{1}{\sqrt{5}}\sum_{j=0}^{4}|jj\rangle\\
|\Psi_{n_{1}1}\rangle & =\left(I\otimes U_{n_{1}1}\right)|\Psi_{00}\rangle,\\
|\Psi_{n_{1}^{\prime}1}\rangle & =\left(I\otimes U_{n_{1}^{\prime}1}\right)|\Psi_{00}\rangle,\\
|\Psi_{n_{2}2}\rangle & =\left(I\otimes U_{n_{2}2}\right)|\Psi_{00}\rangle.\end{align*}
According to Corollary 3, a necessary condition for these four states
to be perfectly distinguishable by one-way LOCC in the direction $A\rightarrow B$,
is that there must exist a vector (normalized) $|\phi\rangle=\sum_{j=0}^{4}\phi_{j}|j\rangle\in\mathcal{H}_{B}$
satisfying the normalization condition \begin{equation}
\sum_{j-0}^{4}|\phi_{j}|^{2}=1\label{example-2:eq:1}\end{equation}
and such that the following four vectors $|\phi\rangle,U_{n_{1}1}|\phi\rangle,U_{n_{1}^{\prime}1}|\phi\rangle,U_{n_{2}2}|\phi\rangle$
are pairwise orthogonal. We now write the orthogonality conditions:\begin{eqnarray}
\langle\phi|U_{n_{1}1}|\phi\rangle=\sum_{j=0}^{4}\omega^{n_{1}j}\phi_{j}\phi_{j\oplus_{5}1}^{*} & = & 0,\label{example-2-eq-2}\\
\langle\phi|U_{n_{1}^{\prime}1}|\phi\rangle=\sum_{j=0}^{4}\omega^{n_{1}^{\prime}j}\phi_{j}\phi_{j\oplus_{5}1}^{*} & = & 0,\label{example-2-eq-3}\\
\langle\phi|U_{n_{1}1}^{\dagger}U_{n_{2}2}|\phi\rangle=\sum_{j=0}^{4}\omega^{(n_{2}-n_{1})j}\phi_{j}\phi_{j\oplus_{5}1}^{*} & = & 0,\label{example-2-eq-4}\\
\langle\phi|U_{n_{1}^{\prime}1}^{\dagger}U_{n_{2}2}|\phi\rangle=\sum_{j=0}^{4}\omega^{(n_{2}-n_{1}^{\prime})j}\phi_{j}\phi_{j\oplus_{5}1}^{*} & = & 0,\label{example-2-eq-5}\\
\langle\phi|U_{n_{2}2}|\phi\rangle=\sum_{j=0}^{4}\omega^{n_{2}j}\phi_{j}\phi_{j\oplus_{5}2}^{*} & = & 0,\label{example-2-eq-6}\\
\langle\phi|U_{n_{1}1}^{\dagger}U_{n_{1}^{\prime}1}|\phi\rangle=\sum_{j=0}^{3}\omega^{(n_{1}^{\prime}-n_{1})j}|\phi_{j}|^{2} & = & 0,\label{example-2-eq-7}\end{eqnarray}
where all the exponents of $\omega=e^{\frac{2\pi i}{5}}$ are taken
to be numbers addition modulo 5. For the set of values $n_{1}=0,n_{1}^{\prime}=3$,
and $n_{2}=2$ (other suitable choices of $n_{1},n_{1}^{\prime},n_{2}$
are also possible) from Eqs.$\,$(\ref{example-2-eq-2})-(\ref{example-2-eq-5}
we see that the vector $\left(\phi_{0}^{*}\phi_{1},\phi_{1}^{*}\phi_{2},\phi_{2}^{*}\phi_{3},\phi_{3}^{*}\phi_{4},\phi_{4}^{*}\phi_{5}\right)\in\mathbb{C}^{5}$
is orthogonal to the set of following four vectors $\left\{ \left(1,1,1,1,1\right),\left(1,\omega^{3},\omega,\omega^{4},\omega^{2}\right),\left(1,\omega^{2},\omega^{4},\omega,\omega^{3}\right),\left(1,\omega^{4},\omega^{3},\omega^{2},\omega\right)\right\} \in\mathbb{C}^{5}.$
Noting that the vector $\left(1,\omega,\omega^{2},\omega^{3},\omega^{4}\right)$
is orthogonal to the previous four vectors, the following identity
\begin{equation}
\left(\phi_{0}^{*}\phi_{1},\phi_{1}^{*}\phi_{2},\phi_{2}^{*}\phi_{3},\phi_{3}^{*}\phi_{4},\phi_{4}^{*}\phi_{5}\right)=\lambda\left(1,\omega,\omega^{2},\omega^{3},\omega^{4}\right)\label{example-2-eq-8}\end{equation}
must be valid for some $\lambda\in\mathbb{C}^{5}$. Proceeding as
in example 1, we need to consider two cases, namely, $\lambda\neq0$
and $\lambda=0$. 

Case 1 ($\lambda\neq0$): This means that $\phi_{j}\neq0$ for all
$j=01,2,3,4$. Using Eq.$\,$(\ref{example-2-eq-8}) in Eq.$\,$(\ref{example-2-eq-6})
we see that \[
\lambda^{*^{2}}\omega^{4}\sum_{j=0}^{4}\frac{1}{|\phi_{j\oplus_{5}1}|^{2}}=0,\]
implying that $\lambda=0$, which is a contradiction. 

Case 2 ($\lambda=0$): Here we need to consider several possibilities
depending upon the values of $\phi_{i}$s. A straightforward but tedious
calculation shows that all the possibilties are ruled out for not
being able to satisfy the orthogonality conditions and Eq.$\,$(\ref{example-2-eq-8})
simultaneously unless $|\phi\rangle$ is a null vector. This therefore
completes the proof. $\square$\\

\emph{Proof of example 3}. As in the proof of the previous example,
we begin with a more general family of five orthogonal states in $6\otimes6$.
We will prove the local indistinguishability in the direction $A\rightarrow B$.
We note that a similar proof holds in the direction $B\rightarrow A$
as well. The states are defined as follows:\begin{eqnarray*}
|\Psi_{00}\rangle & = & \frac{1}{\sqrt{6}}\sum_{j=0}^{5}|jj\rangle,\\
|\Psi_{n_{1}1}\rangle & = & \left(I\otimes U_{n_{1}1}\right)|\Psi_{00}\rangle,\\
|\Psi_{n_{1}^{\prime}1}\rangle & = & \left(I\otimes U_{n_{1}^{\prime}1}\right)|\Psi_{00}\rangle,\\
|\Psi_{n_{2}2}\rangle & = & \left(I\otimes U_{n_{2}2}\right)|\Psi_{00}\rangle,\\
|\Psi_{n_{3}3}\rangle & = & \left(I\otimes U_{n_{3}3}\right)|\Psi_{00}\rangle,\end{eqnarray*}
where, $U_{nm}=\sum_{j=0}^{5}e^{\frac{2n\pi ij}{6}}|j\oplus_{6}m\rangle\langle n|$,
with $n,m=0,1,2,3,4,5$ and $j\oplus_{6}m=\left(j+m\right)\mbox{mod}6$.
Also, we denote $\omega=e^{\frac{2\pi i}{6}}$. 

From Corollary 3, a necessary condition that the above five states
to be perfectly distinguishable by one way LOCC in the direction $A\rightarrow B$
is that there must exist a normalized vector $|\phi\rangle=\sum_{j=0}^{5}\phi_{j}|j\rangle\in\mathbb{C}^{6}$
satisfying the normalization condition \begin{equation}
\sum_{j=0}^{5}|\phi_{j}|^{2}=1\label{ex-3-eq-1}\end{equation}
and such that the following five vectors $|\phi\rangle,U_{n_{1}1}|\phi\rangle,U_{n_{1}^{\prime}1}|\phi\rangle,U_{n_{2}2}|\phi\rangle$
and $U_{n_{3}3}|\phi\rangle$ are pairwise orthogonal. The orthogonality
conditions can be written as, \begin{eqnarray}
\langle\phi|U_{n_{1}1}|\phi\rangle & = & \sum_{j=0}^{5}\omega^{n_{1}j}\phi_{j}\phi_{j\oplus_{6}1}^{*}=0,\label{ex-3-eq-2}\\
\langle\phi|U_{n_{1}^{\prime}1}|\phi\rangle & = & \sum_{j=0}^{5}\omega^{n_{1}^{\prime}j}\phi_{j}\phi_{j\oplus_{6}1}^{*}=0,\label{ex-3-eq-3}\\
\langle\phi|U_{n_{1}1}^{\dagger}U_{n_{2}2}|\phi\rangle & = & \sum_{j=0}^{5}\omega^{\left(n_{2}-n_{1}\right)j}\phi_{j}\phi_{j\oplus_{6}1}^{*}=0,\label{ex-3-eq-4}\\
\langle\phi|U_{n_{1}^{\prime}1}^{\dagger}U_{n_{2}2}|\phi\rangle & = & \sum_{j=0}^{5}\omega^{\left(n_{2}-n_{1}^{\prime}\right)j}\phi_{j}\phi_{j\oplus_{6}1}^{*}=0,\label{ex-3-eq-5}\\
\langle\phi|U_{n_{2}2}^{\dagger}U_{n_{3}3}|\phi\rangle & = & \sum_{j=0}^{5}\omega^{\left(n_{3}-n_{2}\right)j}\phi_{j}\phi_{j\oplus_{6}1}^{*}=0,\label{ex-3-eq-6}\\
\langle\phi|U_{n_{2}2}|\phi\rangle & = & \sum_{j=0}^{5}\omega^{n_{2}j}\phi_{j}\phi_{j\oplus_{6}2}^{*}=0,\label{ex-3-eq-7}\\
\langle\phi|U_{n_{1}1}^{\dagger}U_{n_{3}3}|\phi\rangle & = & \sum_{j=0}^{5}\omega^{\left(n_{3}-n_{1}\right)j}\phi_{j}\phi_{j\oplus_{6}2}^{*}=0,\label{ex-3-eq-8}\\
\langle\phi|U_{n_{1}^{\prime}1}^{\dagger}U_{n_{3}3}|\phi\rangle & = & \sum_{j=0}^{5}\omega^{\left(n_{3}-n_{1}^{\prime}\right)j}\phi_{j}\phi_{j\oplus_{6}2}^{*}=0,\label{ex-3-eq-9}\\
\langle\phi|U_{n_{3}3}|\phi\rangle & = & \sum_{j=0}^{5}\omega^{n_{3}j}\phi_{j}\phi_{j\oplus_{6}3}^{*}=0,\label{ex-3-eq-10}\\
\langle\phi|U_{n_{1}1}^{\dagger}U_{n_{1}^{\prime}1}|\phi\rangle & = & \sum_{j=0}^{5}\omega^{\left(n_{1}^{\prime}-n_{1}\right)j}|\phi_{j}|^{2}=0,\label{ex-3-eq-11}\end{eqnarray}
We choose here $n_{1}=0$, $n_{1}^{\prime}=4$, $n_{2}=1$, and $n_{3}=3$.
Let us note that the proof holds for other suitable choices as well.
From Eqs.$\,$(\ref{ex-3-eq-2}) to (\ref{ex-3-eq-6}) we see that
the vector $\left(\phi_{0}^{*}\phi_{1},\phi_{1}^{*}\phi_{2},\phi_{2}^{*}\phi_{3},\phi_{3}^{*}\phi_{4},\phi_{4}^{*}\phi_{5},\phi_{5}^{*}\phi_{0}\right)\in\mathbb{C}^{6}$
is orthogonal to the incomplete basis $\mathcal{B}\in\mathbb{C}^{6}$
consisting of the following five vectors:\\
$\left(1,\omega^{n_{1}},\omega^{2n_{1}},\omega^{3n_{1}},\omega^{4n_{1}},\omega^{5n_{1}}\right)$,
$\left(1,\omega^{n_{1}^{\prime}},\omega^{2n_{1}^{\prime}},\omega^{3n_{1}^{\prime}},\omega^{4n_{1}^{\prime}},\omega^{5n_{1}^{\prime}}\right)$,
$\left(1,\omega^{\left(n_{2}-n_{1}\right)},\omega^{2\left(n_{2}-n_{1}\right)},\omega^{3\left(n_{2}-n_{1}\right)},\omega^{4\left(n_{2}-n_{1}\right)},\omega^{5\left(n_{2}-n_{1}\right)}\right)$,
$\left(1,\omega^{\left(n_{2}-n_{1}^{\prime}\right)},\omega^{2\left(n_{2}-n_{1}^{\prime}\right)},\omega^{3\left(n_{2}-n_{1}^{\prime}\right)},\omega^{4\left(n_{2}-n_{1}^{\prime}\right)},\omega^{5\left(n_{2}-n_{1}^{\prime}\right)}\right)$,
$\left(1,\omega^{\left(n_{3}-n_{2}\right)},\omega^{2\left(n_{3}-n_{2}\right)},\omega^{3\left(n_{3}-n_{2}\right)},\omega^{4\left(n_{3}-n_{2}\right)},\omega^{5\left(n_{3}-n_{2}\right)}\right)$. 

Note that the vector $\left(1,\omega^{n_{4}},\omega^{2n_{4}},\omega^{3n_{4}},\omega^{4n_{4}},\omega^{5n_{4}}\right)$,
with $n_{4}=5$, completes the above mentioned basis. Therefore we
must have the following relation\begin{equation}
\left(\phi_{0}^{*}\phi_{1},\phi_{1}^{*}\phi_{2},\phi_{2}^{*}\phi_{3},\phi_{3}^{*}\phi_{4},\phi_{4}^{*}\phi_{5},\phi_{5}^{*}\phi_{0}\right)=\lambda\left(1,\omega^{n_{4}},\omega^{2n_{4}},\omega^{3n_{4}},\omega^{4n_{4}},\omega^{5n_{4}}\right)\label{ex-3-eq-12}\end{equation}
 to hold true for some $\lambda\in\mathbb{C}$, and $n_{4}=5$. Now
from Eqs.$\,$(\ref{ex-3-eq-2}) to (\ref{ex-3-eq-6}) we also obtain
that \begin{equation}
\phi_{j}^{*}\phi_{j\oplus1}=\lambda\omega^{n_{4}j}\;\forall j=0,1,2,3,4,5\label{ex-3-eq-13}\end{equation}
We now have to consider two cases, namely when $\lambda\neq0$ and
$\lambda=0$. 

Case 1: Let $\lambda\neq0$. Using Eq.$\,$(\ref{ex-3-eq-13}) into
Eq.$\,$(\ref{ex-3-eq-10}) we see that \begin{equation}
\lambda^{*^{3}}\omega^{3n_{4}}\sum_{j=0}^{5}\frac{\omega^{\left(n_{3}+3n_{4}\right)j}}{|\phi_{j\oplus1}\phi_{j\oplus2}|^{2}}=0\label{ex-3-eq-14}\end{equation}
from which we readily obtain $\lambda=0$ (note that $n_{3}+3n_{4}=0\,\mbox{\mbox{mod}}6)$
which is a contradiction. 

Case 2: Let $\lambda=0$. This gives rise to several subcases that
need to be considered individually. 

Case 2.1: In this case we assume any five elements of the set $\left\{ \phi_{i}:i=0,...,5\right\} $
are zero. Suppose $\phi_{5}\neq0$, and the rest are all zero. The
normalization condition implies that $|\phi_{5}|^{2}=1$. On the other
hand from Eq.$\,$(\ref{ex-3-eq-11}) we see that $\omega^{5\left(n_{1}^{\prime}-n_{1}\right)}|\phi_{5}|^{2}=0$,
thus arriving at a contradiction. Similarly contradictions can be
reached for ther cases as well. 

Case 2.2: Here we assume any four elements of the set $\left\{ \phi_{i}:i=0,...,5\right\} $
are zero. Suppose $\phi_{0}=\phi_{1}=\phi_{2}=\phi_{3}=0$. This clearly
violates Eq.$\,$(\ref{ex-3-eq-13}) and hence this is not possible.
Likewise other cases can also be ruled out. Nevertheless it is instructive
to look at another case in which the proof is slightly more nontrivial.
Suppose $\phi_{0}=\phi_{1}=\phi_{2}=\phi_{4}=0$. Here from Eqs.$\,$(\ref{ex-3-eq-7}),
(\ref{ex-3-eq-8}), and (\ref{ex-3-eq-9}) we obtain\begin{equation}
\phi_{3}\phi_{5}^{*}=0,\label{ex-3-eq-15}\end{equation}
and from Eq.$\,$(\ref{ex-3-eq-11})\begin{equation}
|\phi_{3}|^{2}+\omega^{2\left(n_{1}^{\prime}-n_{1}\right)}|\phi_{5}|^{2}=0,\label{ex-3-eq-16}\end{equation}
 and from the normalization condition we get, \begin{equation}
|\phi_{3}|^{2}+|\phi_{5}|^{2}=1.\label{ex-3-eq-17}\end{equation}
Clearly the above three equations are incompatible. 

Case 2.3: Here we assume any three elements of the set $\left\{ \phi_{i}:i=0,...,5\right\} $
are zero. One can show that all the cases can be ruled out because
contradictions are reached with the orthogonality conditions and/or
Eq.$\,$(\ref{ex-3-eq-13}). We give two instances for better understanding
of the readers. If we take $\phi_{0}=\phi_{1}=\phi_{2}=0$, then this
is clearly in contradiction with Eq.$\,$(\ref{ex-3-eq-13}). Somewhat
more complicated is the proof of the case corresponding to $\phi_{0}=\phi_{2}=\phi_{4}=0$.
From Eqs.$\,$(\ref{ex-3-eq-7}), (\ref{ex-3-eq-8}), and (\ref{ex-3-eq-9}),
and explicitly substituting the values $n_{1}=0$, $n_{1}^{\prime}=4$,
$n_{2}=1$, and $n_{3}=3$, one can show after some simple algebra
that the vector $\left(\phi_{1}^{*}\phi_{3},\phi_{3}^{*}\phi_{5},\phi_{5}^{*}\phi_{1}\right)\in\mathbb{C}^{3}$
is a null vector. That is, \begin{equation}
\left(\phi_{1}^{*}\phi_{3},\phi_{3}^{*}\phi_{5},\phi_{5}^{*}\phi_{1}\right)=\left(0,0,0\right)\label{ex-3-eq-18}\end{equation}
On the other hand Eqs.$\,$(\ref{ex-3-eq-11}) and the normalization
condition give us the following two relations:\begin{eqnarray}
|\phi_{1}|^{2}+\omega^{2(n_{1}^{\prime}-n_{1})}|\phi_{3}|^{2}+\omega^{4(n_{1}^{\prime}-n_{1})}|\phi_{5}|^{2} & = & 0\label{ex-3-eq-19}\\
|\phi_{1}|^{2}+|\phi_{3}|^{2}+|\phi_{5}|^{2} & = & 1\label{ex-3-eq-20}\end{eqnarray}

The above three equations are clearly inconsistent with each other. 

The remaining cases, namely when any two of the elements are zero
and only one element is zero, are easily shown to be ruled out for
they all give rise to conradiction with Eq.$\,$(\ref{ex-3-eq-13}).
This completes the proof. Thus we have shown that the five maximally
entangled states $|\Psi_{00}^{(6)}\rangle,|\Psi_{01}^{(6)}\rangle,|\Psi_{41}^{(6)}\rangle,|\Psi_{12}^{(6)}\rangle,|\Psi_{33}^{(6)}\rangle$
in $6\otimes6$ are not perfectly distinguishable by one way LOCC.
$\square$

\section{Discussions and Conclusions}

We have considered in this work one way local distinguishability of
a set of orthogonal states which are unilaterally transformable. That
is to say, the states can be mapped onto one another by unitary operators
acting on the local Hilbert spaces. We have shown that the one-way
local distinguishability of such states is initimately related to
the question of perfect distinguishability of the corresponding unitary
operators in the local Hilbert space they act upon. In particular,
if the unitary operators cannot be distinguished in their local Hilbert
space but instead are perfectly distinguishable in an extended Hilbert
space, then the set of orthogonal states thus generated are indistinguishable
by one-way LOCC. We then apply these results to distinguish maximally
entangled states by one way LOCC. 

Maximally entangled states, by definition belong to the family of
unilaterally transformable states, although symmetry implies that
maximally entangled states are unilaterally transformable in both
Alice's and Bob's Hilbert spaces. Maximally entangled states are of
considerable importance in quantum information theory and foundations
of quantum mechanics because of their role in quantum communication
primitives like quantum teleportation and superdense coding as well
demonstrating maximal violations of Bell inequalities. Thus local
distinguishability of maximally entangled states has attracted a lot
of attention in the recent years and one of main open questions in
this area is whether a set of $N\leq d$ orthogonal maximally entangled
states in $d\otimes d$ can be perfectly distinguished by LOCC for
all $d\geq4$. 

To help answer this question we have established an one-to-one correspondence
between one-way LOCC distinguishability of a set of orthogonal quantum
states and distinguishability of the local unitary operators which
generate such a set. With the help of this correspondence we have
been able to show that there are sets of $N\leq d$ maximally entangled
states in $d\otimes d$ for $d=4,5,6$ such that these states cannot
be perfectly distinguished by one way LOCC alone. This provides a
strong evidence in support of the conjecture that such sets of states
indeed exist. Very recently in \cite{Yu-Duan2011} a set of four maximally
entangled states in $4\otimes4$ are presented that are not perfectly
distinguishable by PPT operations, and therefore by LOCC but the question
in higher dimensions remain open. We conjecture that these examples
are potentially strong candidates to establish that any $N$ maximally
entangled states in $d\otimes d$ may not be perfectly distinguished
by LOCC even if $N\leq d$. We believe that a resonable way to conclusively
answer this question would be to extend the applicability of the necessary
condition presented in this paper (see Proposition 1 and Corollary
3 for maximally entangled states) to two-way LOCC protocols. 

A very interesting avenue of further research based on the results
presented here would be to extend these results in multipartite systems.
For multipartite systems, unless for very special cases, the extension
is not straightforward and generally gives rise to complex scenarios.
To illustrate let us consider the simplest multipartite scenario consisting
of three parties Alice, Bob and Charlie. Assume that the set of states
are being generated by applying unitary operations on some standard
state, either on the local Hilbert space of Bob or Charlie or both.
Then a straightforward generalization of the bipartite case now gives
rise several independent cases corresponding to the following forms
of unitary operations: (a) $\left\{ I\otimes U_{i}\otimes I\right\} $,
(b) $\left\{ I\otimes I\otimes V_{j}\right\} $, and (c) $\left\{ I\otimes U_{i}\otimes V_{j}\right\} $.
The interesting cases are when the unitary operators $\left\{ U_{i}\right\} $
and $\left\{ V_{j}\right\} $ are not perfectly distinguishable on
the local Hilbert spaces they act upon, and instead can be perfectly
distinguished in an extended tensor product space. 

In the first two cases it is possible to obtain results similar to
that obtained in this work with respect to the following one way LOCC
in the directions $A\rightarrow C\rightarrow B$ for case (a), and
$A\rightarrow B\rightarrow C$ for case (b). On the other hand case
(c) merits careful consideration, and it is not obvious at all how
the results in this paper could be generalized to include such cases.
Thus for a general multipartite system consisting of say, N subsystems,
our results can be applied when the states can be mapped onto one
another by applying local unitaries only one one subsystem. For more
complex scenarios that involve unitaries mapping the states onto one
another by acting on two or more subsystems would call for further
research and beyond the scope of this paper. 

Finally we would like to mention that quantum cryptography primitives
like both classical and quantum data hiding, secret sharing protocols
\cite{Markham-Sanders-2008,Eggeling2002,Terhal2001,DiVincenzo2002}
make use of the fact that it is not possible to perfectly determine
the state of a quantum system even though it was prepared in one of
several orthogonal states. In this paper several examples of locally
one-way indistinguishable minimal (possibly) sets of maximally entangled
states are presented with the property that they are unilaterally
transformable. It is conceivable that these states with their very
special properties may find applications in developing new protocols
for secret sharing and data hiding. 
\begin{acknowledgments}
SG kindly acknowledges the hospitality of Bose Institute, Kolkata
and ISI, Kolkata during his visits in 2010 when part of the work was
completed. \end{acknowledgments}

\end{document}